\documentclass[10pt]{iopart}

\usepackage{graphicx}
\usepackage{amsmath}

\usepackage{hyperref}
\usepackage{cite}
\usepackage{cleveref}
\hypersetup{
    colorlinks=true,
    linkcolor=blue,     
    citecolor=blue,      
    urlcolor=blue,      
}

\begin{document}

\title[Interplay of losses for the frequency- and temperature-dependent quality factor of superconducting resonators]{Interplay of coupling, residual, and quasiparticle losses for the frequency- and temperature-dependent quality factor of superconducting resonators}

\author{Elies~Ben~Achour, Cenk~Beydeda, Gabriele~Untereiner, Martin~Dressel, Marc~Scheffler}

\address{1.~Physikalisches~Institut, Universit\"at~Stuttgart, Pfaffenwaldring~57, 70569~Stuttgart, Germany}
\ead{marc.scheffler@pi1.physik.uni-stuttgart.de}
\vspace{10pt}
\begin{indented}
\item[] \today
\end{indented}

\begin{abstract}
The overall, loaded quality factor $Q_\mathrm{L}$ quantifies the loss of energy stored in a resonator. Here we discuss on general grounds how $Q_\mathrm{L}$ of a planar microwave resonator made of a conventional superconductor should depend on temperature and frequency. We consider contributions to $Q_\mathrm{L}$ due to dissipation by thermal quasiparticles ($Q_\mathrm{QP}$), due to residual dissipation ($Q_\mathrm{Res}$), and due to coupling ($Q_\mathrm{C}$).
We present experimental data obtained with superconducting stripline resonators fabricated from lead (Pb), with different center conductor widths and different coupling gaps. We probe the resonators at various harmonics between 0.7\,GHz and 6\,GHz and at temperatures between 1.5\,K and 7\,K. We find a strongly frequency- and temperature-dependent $Q_\mathrm{L}$, which we can describe by a lumped-element model. For certain resonators at lowest temperatures we observe a maximum in the frequency-dependent $Q_\mathrm{L}$ when $Q_\mathrm{Res}$ and $Q_\mathrm{C}$ match, and here the measured $Q_\mathrm{L}$ can exceed $2 \times 10^{5}$.

\end{abstract}

%
\vspace{2pc}
\ioptwocol

\section{Introduction}

Planar superconducting microwave resonators play an important role in basic and applied research \cite{Goppl,Zmuidzinas_2012}. One prominent direction is coupling GHz photons to solid-state quantum bits for fundamental quantum optics studies and for applications in quantum computing \cite{Wallraff_2004, Frunzio_2005}. Another one employs superconducting resonators as sensitive detectors for photons or particles \cite{Day_2003}. Most popular resonator designs are either based on coplanar waveguides (CPWs) or on lumped element circuits; in both cases the microstructure fabrication requires just a single superconducting layer. Traditionally, a few well-established superconductors such as Al, Nb, or NbN are used for this purpose. More recently, there is increased interest in a variety of other candidate superconductors. One motivation is tuning the superconducting energy gap for low-energy photon detection \cite{Leduc_2010}. Another is employing high kinetic inductance, as found in strongly disordered or granular superconductors \cite{Moshe_2020}, for novel quantum circuitry \cite{Samkharadze_2016, Rotzinger_2017, Gruenhaupt_2018}. Yet another motivation is mitigating material-induced noise in quantum devices \cite{Megrant_2012, McRae_2020}. Towards this goal, residual absorption in the bulk as well as at surfaces and interfaces has to be reduced, and here alternative superconductors such as Ta have been investigated \cite{Crawley_2023}. For all the mentioned applications, superconducting microwave resonators are usually operated at mK temperatures, much lower than the critical temperature \mbox{$T_\mathrm{c}$}.

The strong interest to improve superconductors for various GHz applications converges with another well-established research direction: planar microwave resonators are used to study fundamental electrodynamic properties of various material classes, including superconductors \cite{DiIorio_1988, Oates_1991,  Scheffler_2013, Hafner_2014, JavaheriRahim_2016, Engl_2020}. From such a microwave spectroscopy perspective, one is usually interested in probing the bulk response of a material of interest as function of temperature \cite{Langley_1991, Salluzzo_2000, Driessen_2012, Zemlicka_2015}, ideally also as function of frequency \cite{Oates_1991, Hafner_2014, Feller_2002, Truncik_2013, Thiemann_2018_2}, and often as a function of magnetic field \cite{Revenaz_1994, Powell_1998, Scheffler_2013, Ghirri_2015, Velluire_2023}. For such experiments, the studied material could either constitute a core element of the resonator \cite{Ghigo_2004, Levy-Bertrand_2019} or act as a perturbation to a well-characterized resonator \cite{Scheffler_2015, Ghigo_2017, Ebensperger_2019, Ghigo2023}.

The goal of the present study is bringing together concepts of superconducting planar resonators from quantum circuitry and microwave spectroscopy: we characterize multi-mode resonators towards their limits in quality factor imposed by coupling, thermally induced quasiparticles, and residual losses. This might offer new perspectives in several regards: (1) We employ superconducting stripline (triplate) resonators. These are more difficult to fabricate than e.g.\ CPW resonators, but they have very low radiative losses and are convenient if one wants to probe a bulk sample of a conductor \cite{DiIorio_1988, Scheffler_2012, Scheffler_2013, Hafner_2014}. (2) The resonators are made of lead (Pb), which is a conventional superconductor with substantial \mbox{$T_\mathrm{c}=7.2\,$K}, has low kinetic inductance, and can be fabricated easily as thin films via thermal evaporation. The low-energy properties of superconducting Pb have been studied carefully and act as benchmarks for the fundamentals of BCS theory \cite{Richards_1960, Giaever_1960, Palmer_1968, Klein_1994, Ruby_2015, Gozlinski_2023}. Furthermore, early microwave studies indicated rather low residual losses \cite{Pierce_1973, Judish_1977, Thakoor_1986}, and for spectroscopic studies in magnetic field the \mbox{type-I} superconductivity of Pb with highest critical field of 80\,mT can be helpful \cite{Ebensperger_2016, Thiemann_2018titan}. (3) We explicitly address the frequency dependence \cite{DiIorio_1988, Hafner_2014, Beydeda_2023}, which is highly relevant for spectroscopy, and we focus on the range from 0.7\,GHz to 6\,GHz, which overlaps with the conventional operating range of quantum circuits and detectors. (4) We cover temperatures from 1.5\,K to 7\,K, i.e.\ from temperatures where thermally excited quasiparticles can be neglected in terms of losses until they fully dominate losses close to \mbox{$T_\mathrm{c}$}.

\section{Temperature and Frequency Regimes of $Q_\textrm{L}$}\label{sec:regimes}

The frequency-dependent response of a resonator is often described with the Lorentz model \cite{Pozar_2011}
\begin{equation}
y(f) = \frac{A}{f-f_0 + \mathrm{i}f_\mathrm{b}/2}
\end{equation}
where $A$ is the amplitude, $f$ the frequency, $f_0$ the resonance frequency and $f_\mathrm{b}$ the full width at half maximum (FWHM). For the superconducting transmission line resonators used in this work, $y=S_{21}$ is the forward scattering coefficient. The loaded quality factor
\begin{equation}
    Q_\mathrm{L} = \frac{f_0}{f_\mathrm{b}}
\end{equation}
can be understood as the ratio of energy stored to the energy lost per cycle in a resonator. $Q_\mathrm{L}$ can be straightforwardly extracted from a measurement. 
Losses in a resonator add up linearly in $f_\mathrm{b}$, but it is common to subdivide the loaded quality factor $Q_\mathrm{L}$ into the different loss mechanisms that contribute to $Q_\mathrm{L}$. 
For our case of a superconducting stripline resonator, we distinguish losses through the coupling $Q_\mathrm{C}$, losses due to thermally excited quasiparticles in the superconductor $Q_\mathrm{QP}$, and residual losses $Q_\mathrm{Res}$:
\begin{equation}\label{eq:Q_L paralell in part1}
    \frac{1}{Q_\mathrm{L}} = \frac{1}{Q_\mathrm{C}} + \frac{1}{Q_\mathrm{QP}} + \frac{1}{Q_\mathrm{Res}}.
\end{equation}
Here we assume that dielectric losses in the sapphire substrate and radiation losses in a stripline geometry are negligible. We also neglect the pair-breaking contribution to $Q_\mathrm{L}$ since we probe the resonators at frequencies much lower than the energy gap $hf\ll2\Delta_0$.

\begin{figure}[tbp]
	\centering
	\includegraphics[width=\linewidth]{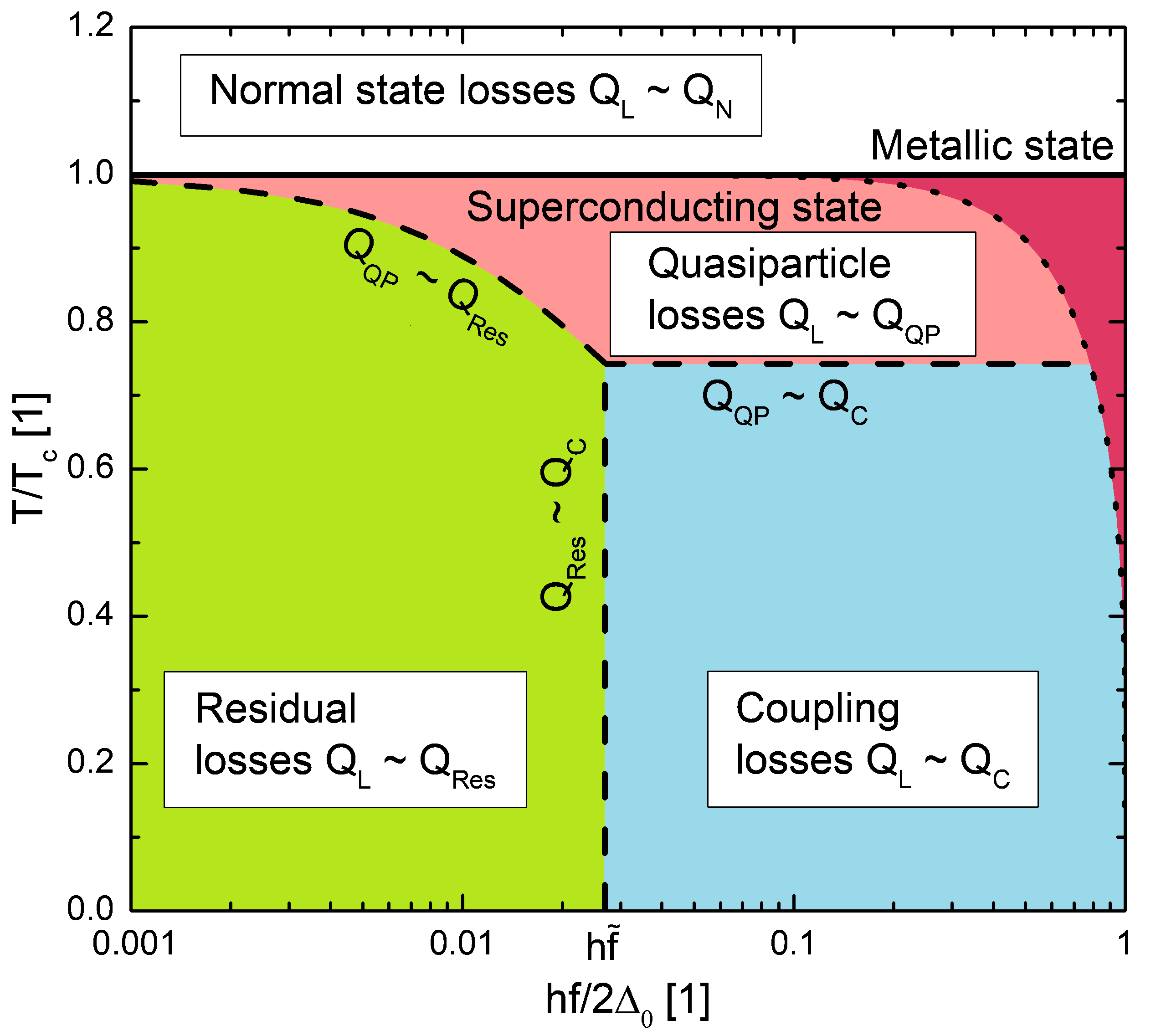}
	\caption{
    Schematic regimes in temperature $T$ (normalized to the critical temperature \mbox{$T_\mathrm{c}$}) and frequency $f$ (normalized to the superconducting energy gap $2\Delta_0$ at zero temperature, with $h$ Planck's constant) concerning different loss mechanisms in a superconducting resonator. 
    The loaded quality factor $Q_\mathrm{L}$ in the superconducting state is dominated by residual losses (quantified by  $Q_\mathrm{Res}$), coupling losses ($Q_\mathrm{C}$), or thermally excited quasiparticles ($Q_\mathrm{QP}$), while above \mbox{$T_\mathrm{c}$} losses due to normal-conductive electrons ($Q_\mathrm{N}$) dominate.
    The dotted line denotes the temperature dependence of the energy gap $2\Delta(T)$, which introduces an additional pair-breaking contribution to $Q_\mathrm{L}$ for frequencies $f > 2\Delta(T)/h$.
    The cross-over frequency $\Tilde{f}$ indicates where the dominant source of losses changes from residual to coupling losses. For details on the shape of the dashed line for $Q_\mathrm{QP}\sim Q_\mathrm{Res}$, see Appendix A.}
	\label{fig:phaseplot}
\end{figure}

In figure \ref{fig:phaseplot} it can be seen which loss mechanism is expected to dominate in such a superconducting resonator for given temperature $T$ and frequency $f$.
At temperatures  close to $T_\mathrm{c}$ the superconducting resonator performance is governed by absorption due to the substantial density of thermally excited quasiparticles  ($Q_\mathrm{QP} \propto 1/f$). Upon further cooling, the loss mechanism changes depending on the frequency $f$. For low frequencies, the resonator performance is limited by residual losses ($Q_\mathrm{Res}\propto f$), which persist even at lowest temperatures. 
Those residual losses result from imperfections in the resonator geometry, or from residual non-superconducting quasiparticles that are not covered by the Mattis-Bardeen formalism \cite{Mattis1958} to describe optical absorption of conventional superconductors based on the BCS theory. For sufficiently high frequency the losses are determined by the resonator coupling ($Q_\mathrm{C} \propto 1/f$), because the impedance of the resonator coupling gap decreases resulting in a stronger leakage through the gap.

The highest quality factor in such a resonator is achieved by a trade-off between residual losses and coupling losses, for a frequency $\Tilde{f}$ at which $Q_{\mathrm{Res}}\sim Q_{\mathrm{C}}$. It is possible to probe this trade-off experimentally by employing multi-mode resonators at several resonance frequencies at lowest temperature.

\begin{figure}[btp]
    \centering
    \includegraphics[width=0.5\textwidth]{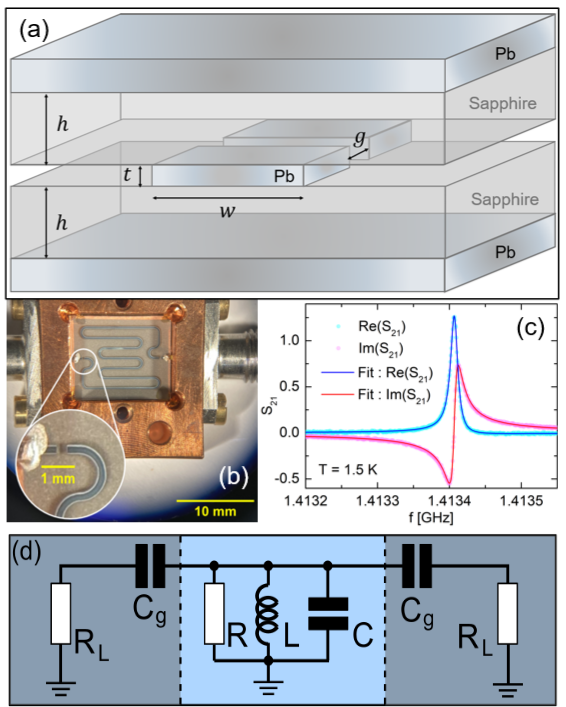}
    \caption{(a) Scheme of a stripline transmission line as used for resonators, with the definition of relevant geometrical parameters. (b) Photograph of the copper box that contains the resonator without the upper sapphire substrate and ground plane. The gray bent line is the center conductor. As shown in the zoomed zone, the center conductor is gapped at two locations such that the center conductor has a finite length $l$. The length $g$ of the gap in the inner conductor is here 300$\,\mu$m. (c) Complex components of $S_{21}$ for resonator 4 around the second resonant mode at $T=1.5\,K$. The solid lines are the results of the fit of the data points using equation~(\ref{eq:fitting function}). The amplitude of the signal exceed unity because an amplifier was used to raise the magnitude of the output signal. (d) Lumped-element circuitry used to model the resonances: the actual resonator are the elements in the blue-shaded middle section, whereas the gray-shaded sections represent the coupling to the environment.}
    \label{fig:exp method}
\end{figure}

\section{Experimental Techniques}

\begin{table*}[tbp]
    \centering
    \begin{tabular}{|c||c|c|c|c|c|c|c|c|c|}
    \hline
         N$^\circ$ & Type & $l$ (cm) & $h$ ($\mu$m) & $w$ ($\mu$m) & $g$ ($\mu$m) & $f_{0, \textrm{des}}$ (GHz) & $f_{0, \textrm{exp}}$ (GHz) & $Q_\mathrm{L}\cdot 10^{-3}$ & $C_\mathrm{g}$ (fF) \\
    \hline
     1 & A & 5 & 430 & 155 & 60 & 0.95 & 0.87 & 15.0 & 17.8 \\
     2 & A & 5 & 430 & 155 & 150 & 0.95 & 0.89 & 87.0 & 7.5 \\
     3 & A & 5 & 430 & 155 & 300 & 0.95 & 0.95 & 90.3 & 6.3 \\
     \hline
     4 & B & 6 & 127 & 45 & 30 & 0.79 & 0.71 & 115.7 & 6.6 \\
     5 & B & 6 & 127 & 45 & 100 & 0.79 & 0.72 & 206.9 & 4.0 \\
     \hline
    \end{tabular}
    \caption{
Overview of the five studied resonators: Relevant geometric parameters as presented in figure \ref{fig:exp method} are resonator length $l$, height $h$ of the dielectric between center conductor and ground planes, and width $w$ of the center conductor. The height $t=1\,\mu$m of the center conductor is the same for all resonators. The length of the gaps $g$ in the center conductor determines the coupling capacitance $C_\mathrm{g}$, which we determine from the experimental data at base temperature. The resonance frequency of the fundamental mode is designed as $f_{0, \textrm{des}}$ and found experimentally as $f_{0, \textrm{exp}}$ at base temperature of 1.5\,K. The observed loaded quality factor $Q_\mathrm{L}$ at base temperature is given for the second resonance mode at 1.7\,GHz and 1.4\,GHz, respectively, for type A and type B geometry.
}
    \label{tab:geometry_parameters}
\end{table*}

In our setup a resonator is a transmission line of finite length $l$ made of Pb. The length of the transmission line results in a discrete set of resonance frequencies 
\begin{equation}
    f_n=\frac{n c}{2l \sqrt{\varepsilon_\mathrm{r}}}, \ n \ge 1
\end{equation}
with a Lorentzian shape in transmission, where $\varepsilon_\mathrm{r}$ is the permittivity of the sapphire dielectric and $c$ the speed of light in vacuum. 
Here we have neglected the contribution of the kinetic inductance of the Pb superconductor.
A measurement of the broadband transmission spectrum of this resonator returns equidistantly spaced transmission peaks of several harmonic modes.
Our experimental setup imposes resonator lengths of a few centimeters, leading to fundamental modes in the microwave range around 1\,GHz. The meander-shaped resonator structure can be seen in figure \ref{fig:exp method}(b).
The impedance mismatch due to the curvature of the conductor is negligible if the radius of curvature is at least three times larger than the width $w$ of the inner conductor \cite{Cohn}.

In figure \ref{fig:exp method}(a) the cross-section of the transmission line can be seen. We use a stripline geometry with an inner conductor of width $w$ and thickness $t$ that is sandwiched between two ground plates where the lower ground plate is 6\,$\mu$m and the upper ground plate 1\,$\mu$m thick.
Both ground plates share the same electrical potential, the center conductor is electrically isolated from the ground plates by two layers of sapphire with thickness $h$.
The center conductor is fabricated by evaporating Pb onto sapphire using a mechanical shadow mask, the upper ground plate is also evaporated on sapphire whereas the lower ground plate consists of a Pb foil.
The center conductor has two gaps with length $g$, as it can be seen in figure \ref{fig:exp method}(b).
The coupling via the gap can be described effectively by a capacitance $C_\mathrm{g}$, which should depend on the geometry of the resonator. 
The resonator structure is mounted in a brass or copper box that shields the resonator and enables electric contact between the two ground plates. The metal box contains two SMA or 1.85\,mm connectors that are connected to the stripline center conductor with silver paste. 

For this study, we use five stripline resonators with different geometry parameters, as listed in table \ref{tab:geometry_parameters}, in order to vary the coupling capacitance value $C_\mathrm{g}$. 
These parameters are chosen such that the length of the coupling gap $g$ varies over a wide range. However, it is always kept smaller than the thickness $h$ of the substrate, $g<h$, to prevent coupling via the ground plates. 
Therefore, we use two types of geometries with different $h$ : type A with $h=$ 430$\,\mu$m and type B with $h=$127$\,\mu$m. Furthermore, since the characteristic impedance $Z_0$ of the transmission line is fully determined by the ratio $w/h$ of the center conductor width and the substrate thickness \cite{Collins}, the width $w$ is adjusted for each resonator type to match the characteristic impedance to 50\,$\Omega$.

The resonators are placed in a cryostat with base temperature below 1.5\,K \cite{Rausch_2018} and connected to a two-port vector network analyzer (VNA) via coaxial cables that measures the spectrum of the resonators from 300\,kHz to 20\,GHz. Typical applied microwave powers are around -40\,dBm.
We measure at temperatures $1.5\,\mathrm{K}<T<8\,$K and record for each resonator mode the microwave transmission spectrum $S_{21}$ with high resolution. 
The measured complex components of $S_{21}$ at resonance are fitted for each temperature, allowing to extract the resonant frequencies and the corresponding FWHM. We use the fitting function 
\begin{equation}\label{eq:fitting function}
S_{21}^{\mathrm{fit}}(f)=e^{2\pi i f \tau}\left( \frac{A}{f_\mathrm{b}/2+i(f-f_0)} + a + b(f-f_0)\right).
\end{equation}
The complex parameters $a$ and $b$ model a linear background in the signal. The real coefficient $\tau$ accounts for a phase shift between the real and the imaginary parts of the signal, which often occurs, due to some phase delay induced by the cables.
The inset in figure \ref{fig:exp method}(c) presents the result of this fitting procedure for an exemplary resonance.

\section{Analysis}

\begin{figure*}[htbp]
    \centering
    \includegraphics[width=\textwidth]{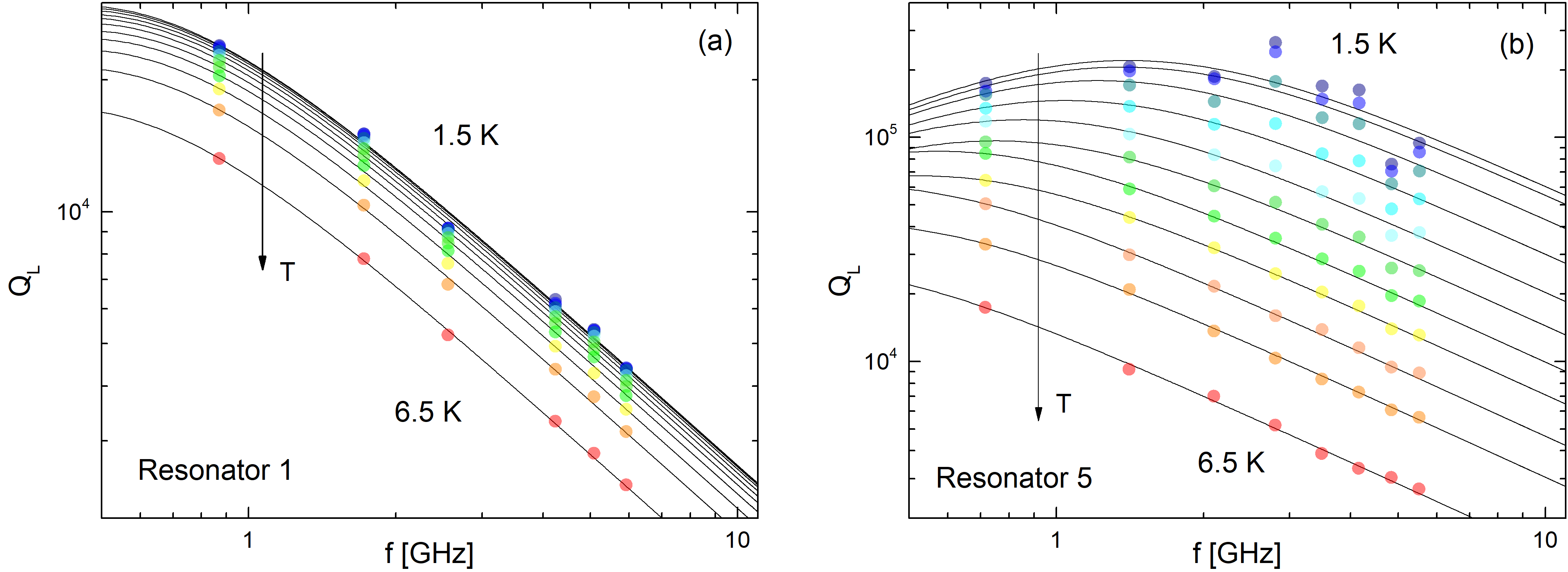}
    \caption{Loaded quality factor $Q_\mathrm{L}$ as a function of frequency $f$ for temperatures from 1.5\,K to 6.5\,K with increment of 0.5\,K for resonator 1 in panel (a) and resonator 5 in panel (b). The continuous lines correspond to the fit of the data points using equation~(\ref{eq:Q_L(f, T)}) for each temperature independently. All resonators of a given type (A or B) have similar frequency and temperature dependence of their quality factor. Both type A and B resonators show the same behavior at high frequency but type B resonators, with thinner substrate, feature a maximum at low frequency (around 1\,GHz for resonator 5).}
    \label{fig:Quality versus freq}
\end{figure*}
The quality factors of the resonance modes can be extracted from the measured spectra for every set temperature. The different resonators span a substantial range of accessible $Q_\mathrm{L}$, as can be seen from table \ref{tab:geometry_parameters}, where the value for $Q_\mathrm{L}$ for the second resonance mode at $T=1.5$\,K is given for each resonator.
Figure \ref{fig:Quality versus freq} shows the frequency dependence of the quality factor at different temperatures for two resonators of different type. These dependences seem to be similar at highest temperatures and frequencies but not for low temperature and frequency, where the data points form a maximum around 1\,GHz for resonator 5. This would suggest that there exist at least two relevant loss mechanisms that exhibit different frequency dependence. 

To investigate this, we use a lumped-element model for the resonators as presented by Göppl \textit{et al.\ }in \cite{Goppl}, and shown in figure \ref{fig:exp method}(d). This model supposes that the transmission line of length $l$ can be modeled as a parallel RLC circuit near resonance and for small losses. It defines an internal quality factor
\begin{equation}\label{eq:Q int}
    Q_{\mathrm{int}}=\omega_nRC
\end{equation}
with resonant frequencies $\omega_n=n\cdot 2\pi f_0 = n(LC)^{-1/2}$. The inductance $L$ and capacitance $C$ can be computed knowing the geometry of the transmission line, and $R$ is a resistance that models the internal losses. 
Assuming symmetric coupling gaps and a low coupling regime ($\omega_nR_\mathrm{L}C_\mathrm{g}\ll 1$) one can quantify a coupling quality factor \cite{Goppl}
\begin{equation}\label{eq:Q coupling}
    Q_{\mathrm{C}}=\frac{C}{2\omega_n R_\mathrm{L} C_\mathrm{g}^2}
\end{equation}
where $R_L=50\,\Omega$ is the impedance imposed by the VNA and the connecting coaxial cables, and $C_\mathrm{g}$ is the coupling capacitance. The loaded quality factor is the parallel association of these two quality factors, such that: 
\begin{equation}\label{eq:Q_L parallel}
    Q_\mathrm{L}^{-1}=Q_{\mathrm{int}}^{-1}+Q_\mathrm{C}^{-1}
\end{equation}
In this model, the only unknown parameters are the coupling capacitance $C_\mathrm{g}$ and the resistance $R$. The latter is inversely proportional to the surface resistance $R_\mathrm{S}$ of Pb, which is a function of frequency and temperature. In our frequency regime, it had been predicted and measured \cite{Dressel2013ElectrodynamicsOM,Hafner_2014} that the surface resistance has the approximate form 
\begin{equation}\label{eq:surface resistance model}
    R_\mathrm{S}(\omega,T)=\rho_\mathrm{S}(T)\omega^2+R_{\mathrm{Res}}(T)
\end{equation}
where $\rho_\mathrm{S}(T)$ quantifies the losses due to thermally excited quasiparticles and would only be a function of temperature. $R_{\mathrm{Res}}$ models the possible residual losses due to Pb itself, due to geometrical defects in the structure, or due to unknown loss mechanisms. Therefore, equations (\ref{eq:Q int})-(\ref{eq:surface resistance model}) would lead to the following form for the loaded quality factor: 
\begin{equation}\label{eq:Q_L(f, T)}
    Q_\mathrm{L}^{-1}=\alpha(T)f+\beta(T)\frac{1}{f}
\end{equation}
with 
\begin{align}
        &\alpha(T)=kC_\mathrm{g}^2+k'\rho_\mathrm{S}(T) \vspace{0.15cm}\label{eq:alpha}\\
         &\beta(T)=k'R_{\mathrm{Res}}(T)\label{eq:beta}
\end{align}
where $k$ and $k'$ are two parameters that can be computed: $k$ accounts for the contribution of the coupling losses to $Q_\mathrm{L}$, and $k'$ quantifies how intrinsic loss mechanisms of the resonator affect $Q_\mathrm{L}$. In this study one only needs to compute $k=4\pi R_\mathrm{L}/C$, where $C$ is obtained using conformal mapping \cite{Collins}. For the two different resonator types we obtain $k(\mathrm{type\,A})=1.2 \times 10^{14} \, \Omega$F$^{-1}$ and $k(\mathrm{type\,B})=1.0 \times 10^{14} \, \Omega$F$^{-1}$, respectively. The final form for the loaded quality factor in equations (\ref{eq:Q_L(f, T)})-(\ref{eq:beta}) leads to equation~(\ref{eq:Q_L paralell in part1}) where $Q_{\mathrm{C}}^{-1}=kC_\mathrm{g}^2f$, $Q_\mathrm{QP}^{-1}=k'\rho_\mathrm{S}f$ and \mbox{$Q_\mathrm{Res}^{-1}=k'R_\mathrm{Res}/f$}. The highest quality factor according to equation~(\ref{eq:Q_L(f, T)}) occurs for every temperature at a cross-over frequency \mbox{$\Tilde{f}(T)=\sqrt{\beta(T)/\alpha(T)}$}.

At high frequency $f>\Tilde{f}$, according to equation~(\ref{eq:Q_L(f, T)}), one would have $Q_\mathrm{L}(f>\Tilde{f})\propto f^{-1}$. In this regime, the losses are dominated both by the quasiparticles and the external coupling, where the ratio between the two depends on the temperature. 
Close to $T_\mathrm{c}$, the large number of quasiparticle excitations results in a rise of the surface resistance which lowers $Q_\mathrm{QP}$, therefore the quasiparticle loss dominates the loaded quality factor $Q_\mathrm{L}\sim Q_\mathrm{QP}$. For low temperatures (and again $f>\Tilde{f}$), coupling losses dominate $Q_\mathrm{L}\sim Q_\mathrm{C}$.
In principle, the coupling loss should be temperature independent. Thus, for determining the coupling capacitance, one will suppose that \mbox{$\rho_\mathrm{S}(T)\to 0$} at base temperature of 1.5\,K, such that $Q_\mathrm{L}$ is given by $C_\mathrm{g}$ in this regime. 
The crossover from \mbox{$Q_\mathrm{L}\sim Q_\mathrm{QP}$} to $Q_\mathrm{L}\sim Q_\mathrm{C}$ upon cooling at high frequency cannot be extracted from the frequency dependence of $Q_\mathrm{L}$, since both $Q_\mathrm{QP}$ and $Q_\mathrm{C}$ are anti-proportional to frequency. With respect to figure \ref{fig:phaseplot}, it corresponds to the crossing of the dashed line from the pink zone to the blue one.

\section{Experimental Results}

The transition from the low-frequency dependence $Q_\mathrm{L}(f<\Tilde{f})\propto f$ to the high-frequency dependence \mbox{$Q_\mathrm{L}(f>\Tilde{f})\propto f^{-1}$} is visible for resonator 5 in figure \ref{fig:Quality versus freq} around 1\,GHz, for the lowest temperatures only. The quasiparticle losses contribution is negligible at base temperature. Thus, the relevant loss mechanisms are the residual and the external coupling losses. For sufficiently small coupling losses at low frequency, the residual losses dominate $Q_\mathrm{L}\sim Q_\mathrm{Res}$, resulting in a linear frequency dependence of $Q_\mathrm{L}$. This would suggest that resonator 1 shows stronger coupling than resonator 5, since $\Tilde{f}$ is shifted to lower frequencies (resulting from higher $\alpha\sim C_\mathrm{g}^2$ at low $T$).

\begin{figure}[tbph]
    \centering
    \includegraphics[width=0.45\textwidth]{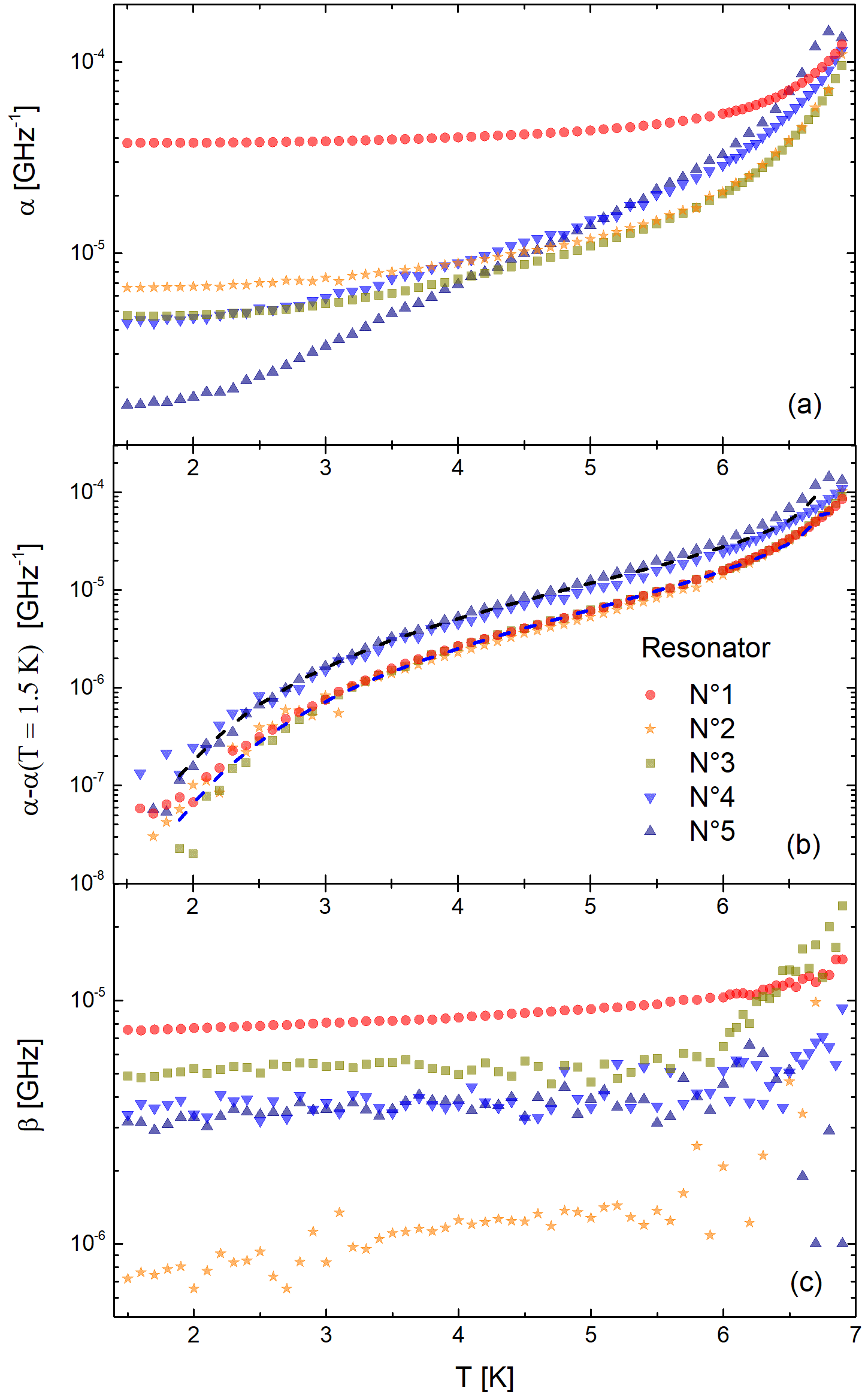}
    \caption{Resulting fit parameters $\alpha$ and $\beta$ from the fit with equation~(\ref{eq:Q_L(f, T)}) to the frequency-dependent $Q_\mathrm{L}$ for each temperature and for the five resonators. The parameter $\alpha$  in (a) accounts for the quasiparticle losses through the temperature-dependent $\rho_\mathrm{S}$ and the temperature-independent coupling losses. One assumes that $\rho_\mathrm{s}\to 0$ at base temperature of 1.5\,K, allowing to estimate the coupling capacitance $C_\mathrm{g}$. 
Panel (b) shows $\alpha(T)-\alpha(T=1.5\,\mathrm{K})$ for the five resonators, which separate into two bundles corresponding to two resonator types. The fits are based on the expected temperature dependence of the surface resistance, see Appendix B for details.
The parameter $\beta$ in (c) is directly proportional to the residual losses modeled by the resistance $R_\mathrm{Res}$.}
    \label{fig:alpha beta}
\end{figure}

For higher temperature, the quasiparticle losses rise ($\alpha$ increases), resulting in $\Tilde{f}$ shifting to lower frequencies.
Thus, for all resonators at high temperatures the cross-over frequency does not appear in our frequency window, since $Q_\mathrm{QP}$ will be always smaller than $Q_\mathrm{Res}$.

Using equation~(\ref{eq:Q_L(f, T)}), the frequency dependence of $Q_\mathrm{L}$ is fitted, with $\alpha$ and $\beta$ as fit parameters.
The result of these fits are plotted as black lines in figure \ref{fig:Quality versus freq}. Each curve corresponds to a fixed temperature and is obtained independently. 
The resulting fit parameters $\alpha$ and $\beta$ are plotted as function of temperature for all five resonators in figure \ref{fig:alpha beta}(a) and \ref{fig:alpha beta}(c), respectively.

These parameters allow us to access different physical quantities of interest. First, assuming that $\rho_\mathrm{S}(T)\to0$ at base temperature, $\alpha(T\to0)$ gives a value of $C_\mathrm{g}$ for each resonator. The results for these measurements are given in table \ref{tab:geometry_parameters}. One can observe here that a smaller cross-section of the inner conductor results in a lower coupling capacitance and that (for a given type of resonator) $C_\mathrm{g}$ is a decreasing function of the gap length $g$. Both of these observations are consistent with the behavior of a planar capacitance with respect to its geometry.
However, for similar ratio $tw/g$, the capacitance is lower for smaller $h$, as the results for resonator 3 and 5 can show. Therefore, the parameter $h$ should play a role for the coupling capacitance, which seems reasonable since its order of magnitude is similar to those for $g$ or $w$. 

A second physical parameter that can be discussed is the temperature dependence of the surface resistance $R_\mathrm{S}$ of Pb, encoded in the quantity $\rho_\mathrm{S}(T)$, see equation~(\ref{eq:surface resistance model}). This quantity seems to reach higher values as the temperature gets closer to $T_\mathrm{c}$. Indeed, a higher density of thermally excited quasiparticles is expected for increasing temperature, and even more so upon approaching $T_\mathrm{c}$ and thus shrinking of the superconducting energy gap. This behavior can be quantitativly studied through the quantity \mbox{$\alpha(T)-\alpha(T=1.5\,K)$}, see figure \ref{fig:alpha beta}(b), which is proportional to $\rho_s(T)$. These experimental data are fitted separately for each resonator type and show good agrement with the prediction of the temperature dependence of $R_s(\omega,T)$ \cite{Dressel2013ElectrodynamicsOM}, see Appendix B for details.

\begin{figure}[tbp]
    \centering
    \includegraphics[width=0.45\textwidth]{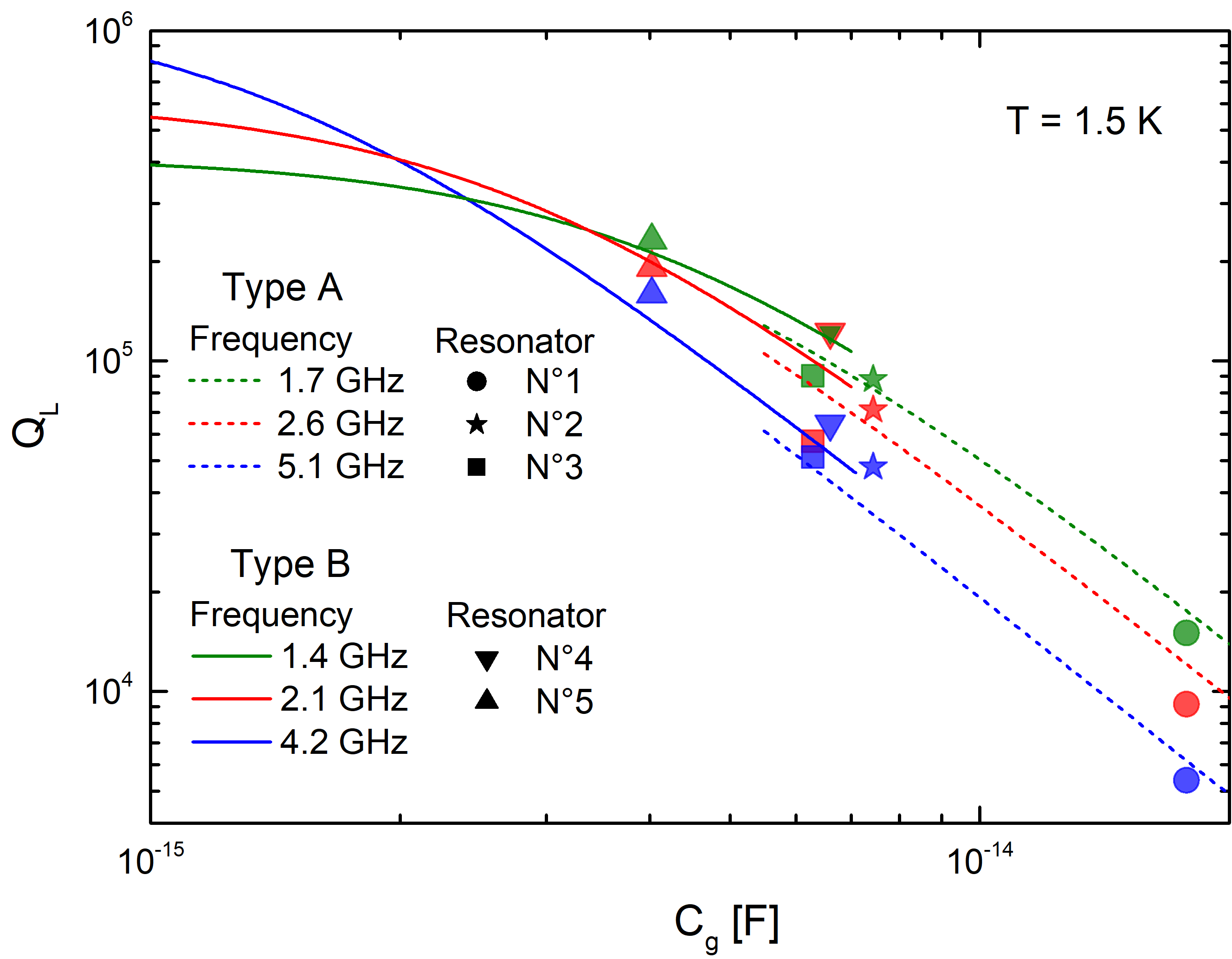}
    \caption{Quality factor dependence with respect to the coupling capacitance for resonance modes 2 (green), 3 (red) and 6 (blue) at $T=$1.5\,K. The solid and dashed lines are obtained from fitting the frequency dependence of $Q_\mathrm{L}$ and by inserting a continuous range for $C_\mathrm{g}$ in equation~(\ref{eq:Q_L(f, T)}). Type A and B have different length, therefore the resonances occur at different frequency.}
    \label{fig:QL of Cg}
\end{figure}

The residual losses measurements suggest that $R_\mathrm{Res}$ is temperature independent, as evident from the almost constant values of $\beta$ for the five resonators in figure \ref{fig:alpha beta}(c).
However, the substantial scattering of the measurements for different resonators could indicate that the measured parameter $R_\mathrm{Res}$ is not intrinsic to Pb but linked to other loss mechanism due, for instance, to the fabrication of the resonators.

In figure \ref{fig:QL of Cg} the total quality factor $Q_\mathrm{L}$ is plotted with respect to the coupling capacitance. The data points in figure \ref{fig:QL of Cg} correspond to modes 2, 3 and 6 at $T=1.5$\,K for each resonator, the solid and dashed lines account for the 2 different lengths $l$ of the resonator types. The lines result from equation~(\ref{eq:Q_L(f, T)}) with varying $C_\mathrm{g}$ and by plugging in a single value of $\beta$ for each resonator type. These values were obtained from averaging the results for $\beta$, at lowest temperature, across the resonators of a given type [$\beta(\mathrm{type\, A})=4.4 \times 10^{-6}$ GHz and $\beta(\mathrm{type\, B})=3.3 \times10^{-6}$ GHz].

One sees that for large $C_\mathrm{g}$ the loaded quality factor follows a $C_\mathrm{g}^{-2}$ trend in agreement with equations (\ref{eq:Q coupling}) and (\ref{eq:Q_L parallel}). For smaller coupling capacitance $Q_\mathrm{L}$ approaches a constant value. Below $C_\mathrm{g}\sim2$\,fF, the contribution of $Q_\mathrm{C}$ to the loaded quality factor $Q_\mathrm{L}$ is negligible. The fact that $Q_\mathrm{L}$ increases with frequency at low capacitance, suggests that in this regime residual losses are dominant. Thus even weaker coupling than in our experiments would be desirable if one wants to achieve the highest possible $Q_\mathrm{L}$ that is only limited by the residual losses.

\section{Discussion}

To finish this study, we give some general remarks. As can be seen in figure \ref{fig:Quality versus freq} for lowest temperatures, the measured quality factor in dependence of frequency does not completely coincide with the fit, but the data are somewhat scattered. This scattering could be well reproduced when the resonators were measured twice in separate cooldowns. This would suggest that the scattering of the quality factor is due to some feature related to the resonator fabrication or its design. As an example, how critical elements of the resonator geometry are challenging to control precisely with our fabrication procedures, the length of the two coupling gaps can differ by 10$\%$ (see the photograph in figure \ref{fig:exp method}(b)), whereas the model assumes perfect symmetry. Finally, as we make use of a lumped model for the coupled resonator, the parameters of the model are taken frequency independent except the resistance $R$. In other words, we assumed that in first order we can attribute all the frequency dependence of $Q_\mathrm{L}$ to the specific features we addressed in this study, although there could possibly exist other mechanisms that could play a role.

\section{Conclusions}

This study addresses how three different loss mechanisms govern the overall loaded quality factor of superconducting resonators: coupling to environment, thermal quasiparticles, and residual losses. For a given resonator design, the dominant loss mechanism depends on temperature and frequency, as illustrated by figure \ref{fig:phaseplot}. We have verified this overall evolution experimentally using five stripline resonators fabricated from superconducting Pb, which differ in geometric dimensions, in particular concerning their coupling capacitance to the connecting microwave lines. In contrast to most studies on planar superconducting resonators, we have explicitly evaluated the frequency dependence by probing several harmonic modes, and there we have demonstrated the transition from $Q_\mathrm{L}$ dominated by residual losses at low frequencies to $Q_\mathrm{L}$ dominated by coupling losses at high frequencies, with a maximum $Q_\mathrm{L}$ at the transition between these ranges.

This work also concerns the potential of Pb as superconducting material for planar microwave resonators.  The electrodynamic properties of superconducting Pb were studied previously at microwave frequencies e.g.\ in the context of three-dimensional cavity resonators \cite{Wilson_1963, Hahn_1968, Pierce_1973, Tsai_2004}, but there are comparably few studies that employ Pb for planar resonators \cite{Scheffler_2012, Hafner_2014, Koepke_2014, Ebensperger_2016}. Our study shows $Q_\mathrm{L}$ around 200.000 for a temperature of 1.5\ K, and following our investigation, this value could be easily exceeded if the coupling would be substantially weakened. This might be difficult to achieve for stripline resonators, and thus future studies might investigate Pb-based coplanar or lumped element resonators. Another open question concerns the power dependence: our work addresses resonator performance at temperatures above 1.5\ K and at intermediate microwave powers around -40\,dBm, as commonly employed in microwave spectroscopy. A different field of applications, quantum circuits, on the other hand typically works at much lower temperatures and  at power levels corresponding to single photons \cite{Zmuidzinas_2012, Megrant_2012}. There the \lq residual losses\rq{} play a different role than in our study, and thus it would be of interest to see how Pb resonators behave in that regime of lower temperatures and powers.

\section*{Acknowledgments}

We thank Ioan Pop, Benjamin Sac\'{e}p\'{e}, and Thibault Charpentier for helpful discussions.
We acknowledge support by the State Graduate Support Program (Landesgraduiertenförderung) of Baden-W\"urttemberg and by the BMBF within the project QSolid (FKZ: 13N16159).

\section*{Appendix A. Calculating the transition between regimes dominated by residual losses and thermal quasiparticles}

The dashed curve in figure \ref{fig:phaseplot} at temperatures close to  $T_\mathrm{c}$ and low frequencies that indicates the transition  between the regime dominated by residual losses and the regime dominated by losses due to thermally excited quasiparticles, $Q_\mathrm{QP}\sim Q_\mathrm{Res}$, was computed via setting \cite{Klein_1993_M}
\begin{equation}
\begin{split}
\frac{Q_\mathrm{QP}}{Q_\mathrm{Res}} &= \frac{Q_\mathrm{QP}}{0.005\cdot Q_\mathrm{N}} = \frac{R_\mathrm{S,\,N}}{0.005\cdot R_\mathrm{S,\,QP}} \\ &= \frac{1}{0.005}\cdot \mathrm{Re}\left[\sqrt{\frac{\hat{\sigma}_\mathrm{QP}}{\hat{\sigma}_\mathrm{n}}}\right],
\end{split}
\end{equation}
where $R_\mathrm{S}$ is the real part of the surface impedance and $\hat{\sigma}$ the optical conductivity, the subscript $\mathrm{N}$ denotes the normal conducting state. The ratio 0.005 between residual losses and normal state losses $Q_\mathrm{N}$ aligns with our results.
The optical conductivity of the quasiparticles $\hat{\sigma}_\mathrm{QP}/\hat{\sigma}_\mathrm{n}$ was computed via the  Zimmermann formula in the dirty limit \cite{Zimmermann_1991}. By using the Zimmermann formula we attribute the quasiparticle losses to low-frequency absorption due to thermally excited quasiparticles and to photon-induced pair-breaking excitations above the superconducting energy gap $\Delta(T)$ of the superconducting condensate. In that sense we assume that there are no quasiparticle losses at lowest temperature since we probe our resonator at frequencies $hf\ll 2\Delta_0$.

\section*{Appendix B. Fitting the temperature-dependent quasiparticle losses}

The fits in figure \ref{fig:alpha beta}(b) use as fit function for the expected temperature dependence of $ R_\mathrm{S}$ \cite{Dressel2013ElectrodynamicsOM}:
\begin{equation}\label{eq:Dressel-Gruner Rs}
    R_\mathrm{S} \propto \frac{(\hbar\omega)^2}{k_\mathrm{B}T}\ln\left\{{\frac{4k_\mathrm{B}T}{\hbar\omega}}\right\}\exp{\left\{-\frac{\Delta(T)}{k_\mathrm{b}T}\right\}}
\end{equation}
with $k_\mathrm{B}$ the Boltzmann constant. For the temperature dependence of the superconducting energy gap $\Delta(T)$ we use the following approximation \cite{Sheahen1966, Sindler_2010}:
\begin{equation}
\frac{\Delta(T)}{\Delta_0} = \sqrt{\cos{\left\{\frac{\pi}{2}\left(\frac{T}{T_\mathrm{c}}\right)^2 \right\}}}
\end{equation}
The fit parameters are  $T_\mathrm{c}=6.7\,$K (both curves) and  $2\Delta_0/k_\mathrm{B}T_\mathrm{c} = 4.5$ (upper curve) and $2\Delta_0/k_\mathrm{B}T_\mathrm{c} = 4.8$ (lower curve). These values exceed the weak-coupling BCS value of $2\Delta_0/k_\mathrm{B}T_\mathrm{c} = 3.53$, which is consistent with previous reports of Pb being a strong coupling superconductor\cite{Carbotte_1990}.

\vspace{0.2cm}
\section*{References}
\bibliographystyle{unsrt} 
\bibliography{refs_2024-12-11}

\end{document}